\documentclass[prb,twocolumn,superscriptaddress]{revtex4-2}

\usepackage[dvipdfmx]{graphicx}
\usepackage{dcolumn}
\usepackage{bm}
\usepackage{color}

\graphicspath{{fig/}}

\begin{document}

\preprint{APS/123-QED}

\title{
Theoretical designing of  multiband Nickelate and Palladate superconductors with $d^{8+\delta}$ configuration 
} 
\author{Naoya Kitamine}
\author{Masayuki Ochi}
\author{Kazuhiko Kuroki}
\affiliation{Department of Physics, Osaka University, 1-1 Machikaneyama-cho, Toyonaka, Osaka, 560-0043, Japan}

\date{\today}

\begin{abstract}
  In a previous study, we proposed  a possibility of high $T_c$ superconductivity in mixed-anion nickelates with $d^{8+\delta}$ electron configuration. The theory was based on the fact that the two-orbital Hubbard model, when all the intra- and interorbital interactions have the same magnitude,  is equivalent to the bilayer Hubbard model, which has been suggested to exhibit high $T_c$ superconductivity.  The energy level offset $\Delta E$ in the two-orbital model is transformed to twice the interlayer hopping in the bilayer model, and hence appropriately large  $\Delta E$ is favorable for superconductivity in the former.  Extending this idea to multiorbital systems, we previously suggested materials with large energy level offset between $d_{x^2-y^2}$ and other $d$ orbitals,  such as Ca$_2$NiO$_2$Cl$_2$,  to be good candidates for high $T_c$ superconductivity, but such materials have not been synthesized to our knowledge. In the present study, we first focus on Sr$_2$NiO$_2$Cl$_2$, which has been synthesized in the past but has small $\Delta E$, and study the effect of applying pressure, which enhances $\Delta E$.  We also study a 4d analogue of Sr$_2$NiO$_2$Cl$_2$, namely, Sr$_2$PdO$_2$X$_2$ ($X=$ Cl, F, H) , in which $\Delta E$ turns out to be large. The analysis using fluctuation exchange approximation suggests possibility of superconductivity in these systems with large $\Delta E$.  We also study the effect of electron doping of these material, which is expected to enhance superconductivity, within the virtual crystal approximation.
\end{abstract}

\pacs{ }
\maketitle
\section{Introduction}

Superconductivity in multiband systems has a long history of theoretical research, starting from seminal studies by Suhl {\it et al.}\cite{Suhl} and Kondo\cite{Kondo}. In multiband systems, the pair scattering between the bands can enhance or even induce superconductivity. In such cases, one might expect that it is most favorable for superconductivity when both of the bands firmly intersect the Fermi level, but recent studies show that in some cases, superconductivity is optimized when one of the bands barely touches the Fermi level, or even lies slightly below (or above) it, namely, when the band is ``incipient''. \cite{DHLee,Hirschfeld,Hirschfeldrev,YBang,YBang2,YBang3,Borisenko,Ding,Kuroki,MaierScalapino2,Matsumoto,Ogura,OguraDthesis,Matsumoto2,DKato,Sakamoto,Kainth,KobayashiAoki,Misumi,Sayyad,Aokireview}.  An example in which a nearly incipient band can strongly enhance superconductivity can be found in the two-leg Hubbard ladder.  In previous studies, one of the present authors have proposed that by doping large amount of electrons in the cuprate two-leg ladder compounds can realize a situation where the top of the bonding band sits close to the Fermi level, while the antibonding band firmly intersects it, and  this can give rise to a very high $T_c$\cite{Kuroki,Sakamoto}. Since doping large amount of electrons in cuprate ladder compounds is difficult, an alternative way of realizing a similar situation was proposed in a bilayer Ruddlesden Popper compound Sr$_3$Mo$_2$O$_7$, where Mo $4d_{xz}$ ( and also $4d_{yz}$) orbitals electronically form a ``hidden ladder''\cite{Ogura,OguraDthesis}.

Another example can be found in the bilayer Hubbard model\cite{KA,MaierScalapino,Bulut,Scalettar,Hanke,Santos,Mazin,Kancharla,Bouadim,Fabrizio,Zhai,Maier,Nakata,MaierScalapino2,Matsumoto2,DKato,Kainth,Congjun,KarakuzuMaier}, which can be considered as a two-dimensional analogue of the two-leg Hubbard ladder.  This model consists of two layers of two-dimensional Hubbard models, typically on square lattices, coupled with vertical hoppings $t_\perp$ between the two layers. Various studies have shown that superconductivity is optimized when one of the bands (bonding or antibonding) touches or nearly touches the Fermi level. This is the case where $t_\perp$ is several times larger than the magnitude of the in plane hopping $t$, and the system is close to half-filling. Some studies suggest that the superconducting transition temperature, scaled by $t$, in the optimized case can be higher than that of the single-layer Hubbard model\cite{KA,MaierScalapino,Nakata,MaierScalapino2,Matsumoto2}, which is often adopted as a model for the high $T_c$ cuprates. As a possible way of realizing a nearly half-filled bilayer Hubbard model in actual materials, one of the present authors proposed that a bilayer Ruddlesden-Popper nickelate, La$_3$Ni$_2$O$_7$, may become a high temperature superconductor, provided that some parameters are tuned in an appropriate manner\cite{Nakata}. In fact, quite recently, high $T_c$ superconductivity of $T_c\sim 80$K has been discovered in La$_3$Ni$_2$O$_7$ under pressure\cite{MWang}, followed by experiments confirming the discovery\cite{2307.09865,2307.14819} as well as various theoretical studies\cite{2306.07837,2306.03706,2306.07931,2307.05662,2307.16697,2307.16873,2307.15706,2305.15564,2307.10144,2307.06806,2307.14965,2307.07154,2306.03231,2307.15276,2306.05121, 2308.01176, 2308.06771, 2306.07275} including ours\cite{sakakibarala327}.

\begin{figure}
\includegraphics[width=8cm]{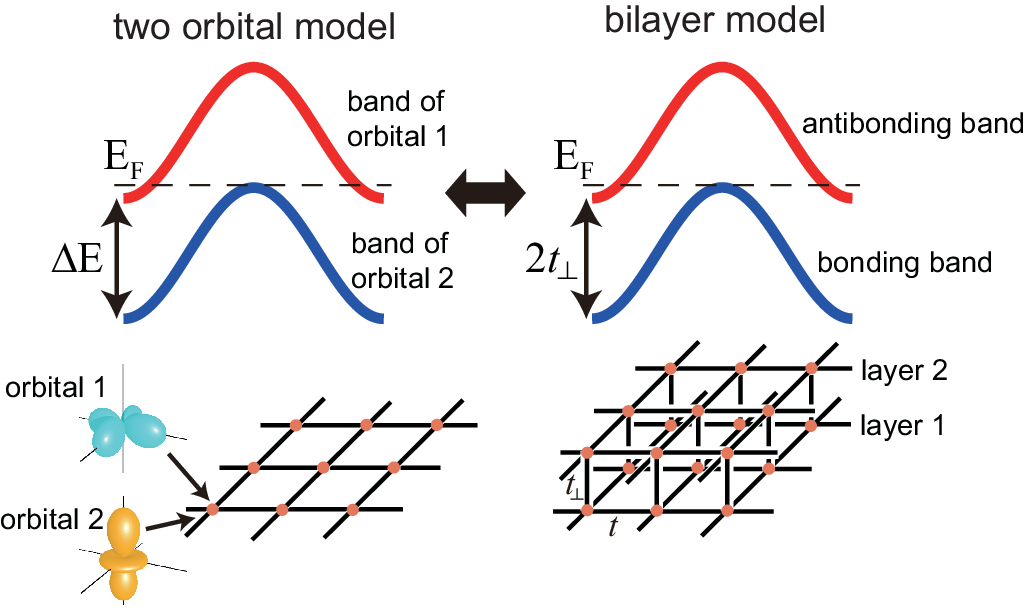}
\caption{Equivalence between two-orbital and bilayer Hubbard models.\label{fig1}}
\end{figure}

Two-leg ladder and bilayer models are examples where bonding and antibonding bands of a {\it single} orbital per site play important roles, but in a previous theoretical study\cite{Yamazaki} on a new type of cuprate superconductor Ba$_2$CuO$_{3+\delta}$\cite{Li}, two of the present authors showed that superconductivity in a two-orbital model can be strongly enhanced when one of the bands (nearly) touches the Fermi level. In a two-orbital model, there are two orbitals per site with one site per unit cell, while in the bilayer and ladder models, there are two sites per unit cell with one orbital per site. In the former, intra- and interorbital electron interactions exist, while in the latter only the on-site repulsion is present. So the two models appear to be quite different, but in Ref.\cite{Yamazaki}, we have attributed the resemblance of the incipient-band enhanced superconductivity to the fact that the two models are mathematically equivalent when the magnitudes of all the intra- and interband interactions are equal\cite{Shinaoka} (Fig.\ref{fig1}).  In this transformation, $2t_\perp$ in the bilayer model corresponds to the orbital level offset $\Delta E$ in the two-orbital model, so that $T_c$ becomes high when $\Delta E$ is appropriately large. In actual materials, the intra- and interorbital interactions have different magnitudes, but we have shown that the similarity between the two models holds to some extent even when the electron interactions in the two-orbital model are taken to be realistic values.

\begin{figure}
\includegraphics[width=8cm]{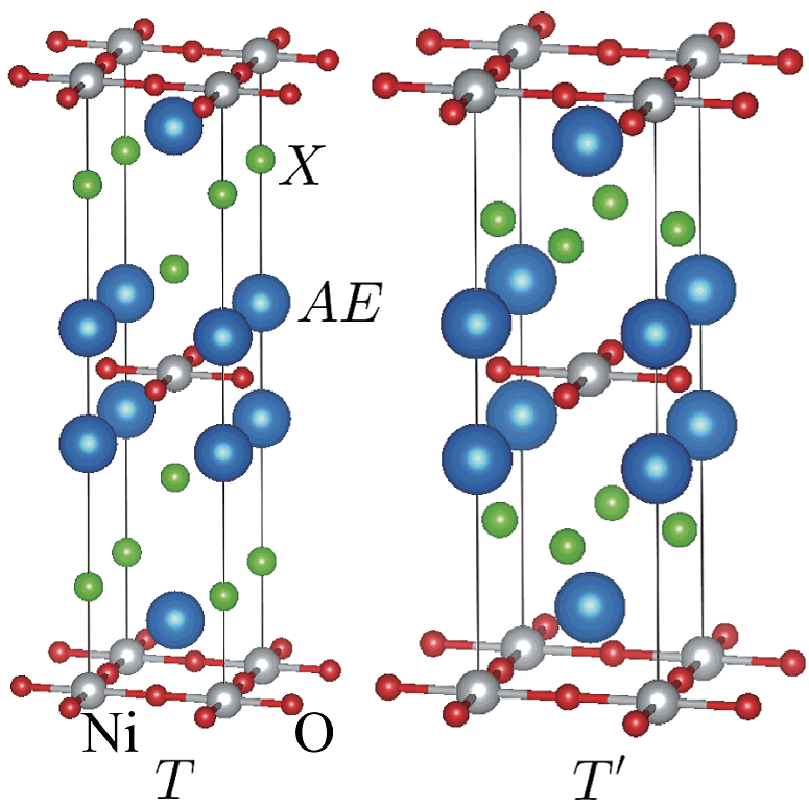}
\caption{The $T-$ and $T'-$ crystal structures of $AE_2$NiO$_2X_2$ with $AE=$Ca,Sr and $X=$Cl, F, H. The figures are drawn by VESTA\cite{VESTA}. \label{fig2}}
\end{figure}

Subsequently, we came up with an idea of realizing a large $\Delta E$ situation in mixed-anion nickelates with nearly $d^8$ electron configuration\cite{Kitamine}. Starting with a $d^8$ nickelate La$_2$NiO$_4$ having a layered perovskite structure, we replace the apical oxygens by chlorines to enhance the crystal field splitting and hence $\Delta E$ between the $d_{x^2-y^2}$ and $d_{3z^2-r^2}$ orbitals. La is substituted by Ca to maintain the $d^8$ electron configuration, and we end up Ca$_2$NiO$_2$Cl$_2$ (with $T$-crystal structure, see Fig.\ref{fig2}), in which the wide $d_{x^2-y^2}$ band intersects the Fermi level, while the other four (relatively narrow) $d$ bands touch or lie slightly below the Fermi level. According to our calculation, superconductivity is optimized by doping certain amount of electrons,  owing to a nearly incipient-band situation 
with a large $\Delta E$.

Here, a problem is that,  to our knowledge,  Ca$_2$NiO$_2$Cl$_2$ has never been synthesized. Although  Sr$_2$NiO$_2$Cl$_2$ has been synthesized\cite{Tsujimoto}, adopting Sr instead of Ca results in a reduction of $\Delta E$ due to the increase of the lattice constant; the increase in the in-plane Ni-O distance results in a suppression of the crystal field splitting. This has motivated our present study of investigating  the effect of applying hydrostatic pressure to  Sr$_2$NiO$_2$Cl$_2$, which is expected to reduce the in-plane lattice constant and hence enhance $\Delta E$. In addition to this, here we also find another possible way to enhance $\Delta E$ and thus superconductivity, that is, considering a 4d analogue of Sr$_2$NiO$_2$Cl$_2$ by replacing Ni by Pd. We note that Ba$_2$NiO$_2$Cl$_2$\cite{Tsujimoto} and Ba$_2$PdO$_2$F$_2$\cite{Baikie}, which are expected to have similar electronic properties with Sr$_2$NiO$_2$Cl$_2$, have also been synthesized.    We construct five orbital models for these materials from first principles calculation, and adopt a combination of  fluctuation exchange (FLEX) approximation and linearized Eliashberg equation, which suggests possibility of superconductivity in these systems with large $\Delta E$.  We also study the effect of electron doping of these material, which enhances superconductivity, within the virtual crystal approximation.
\begin{figure}
\includegraphics[width=8cm]{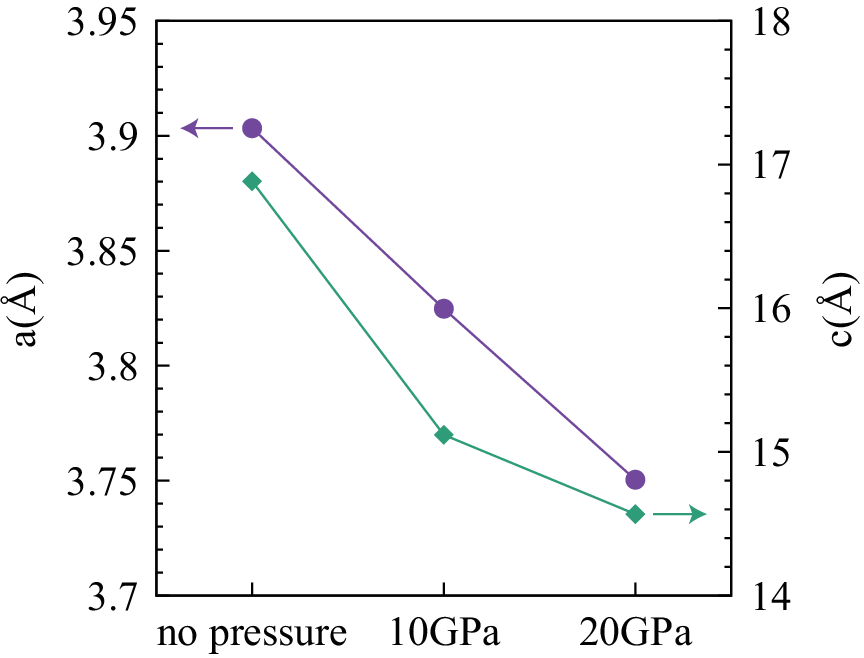}
\caption{The optimized lattice constants of Sr$_2$NiO$_2$Cl$_2$ at ambient or under pressure.\label{fig3}}
\end{figure}   
\begin{figure*}
\includegraphics[width=16cm]{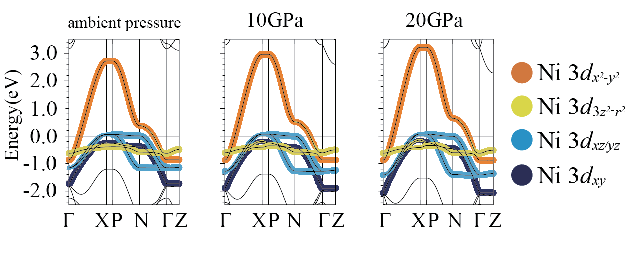}
\caption{The band structure of Sr$_2$NiO$_2$Cl$_2$ at ambient or under pressure.\label{fig4}}
\end{figure*}
\begin{figure}
\includegraphics[width=8cm]{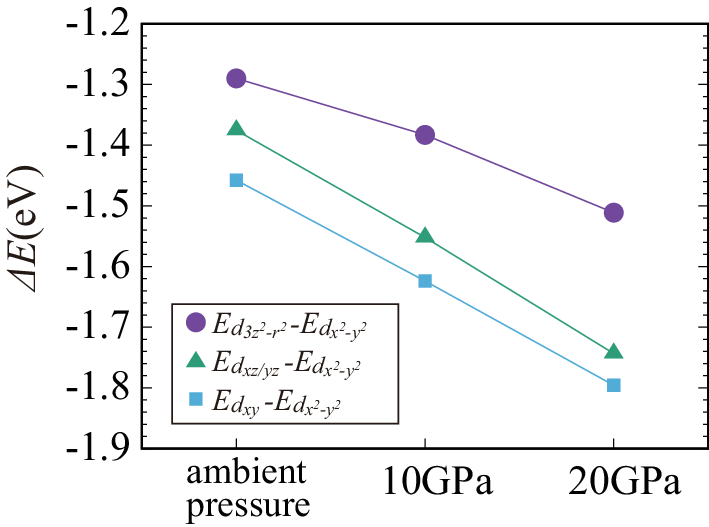}
\caption{The energy level offsets in Sr$_2$NiO$_2$Cl$_2$ between $d_{x^2-y^2}$ and other $d$ orbitals at ambient or under pressure.\label{fig5}}
\end{figure}
\begin{figure}
\includegraphics[width=8cm]{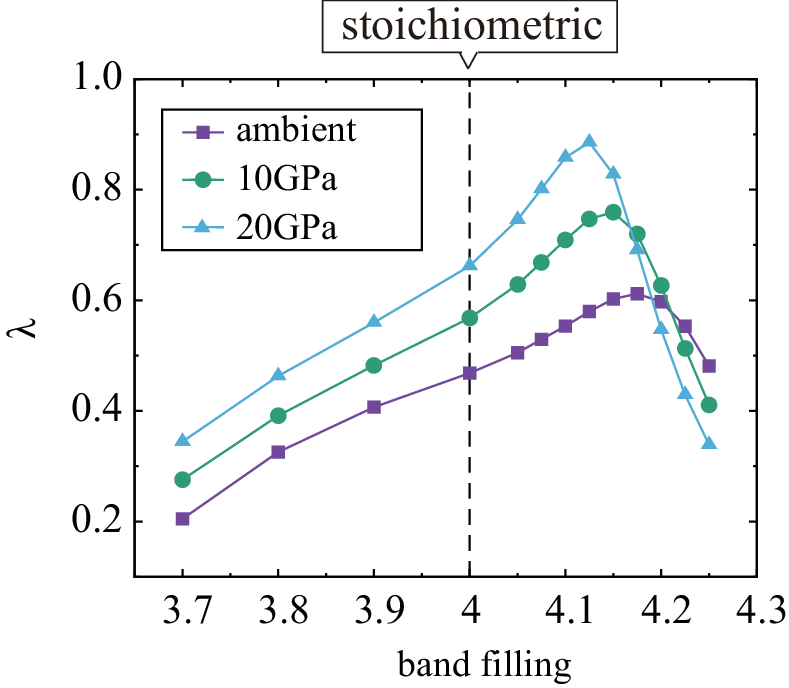}
\caption{The eigenvalue of the linearized Eliashberg equation $\lambda$ at $T=0.01$eV plotted against the band filling at ambient or under pressure.\label{fig6}}
\end{figure}   
\begin{figure}
\includegraphics[width=8cm]{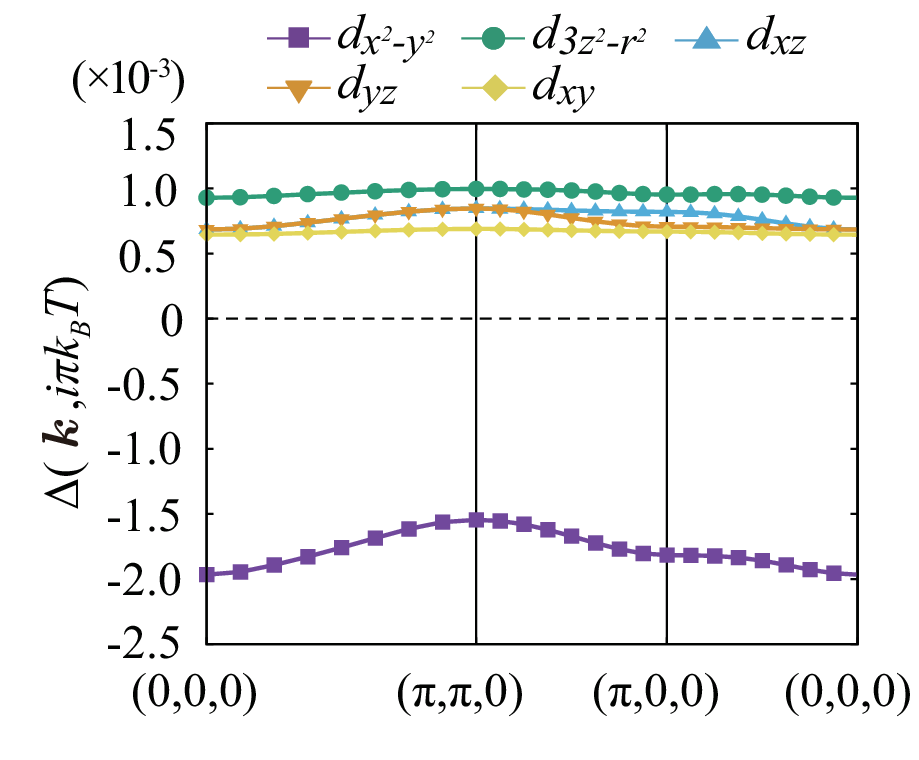}
\caption{The gap function of Sr$_2$NiO$_2$Cl$_2$ in the orbital representation plotted against the wave vector. The parameter values that give the largest $\lambda$ is adopted (pressure of 20GPa and $n=4.12$).\label{fig7}}
\end{figure}

\begin{figure}
\includegraphics[width=8cm]{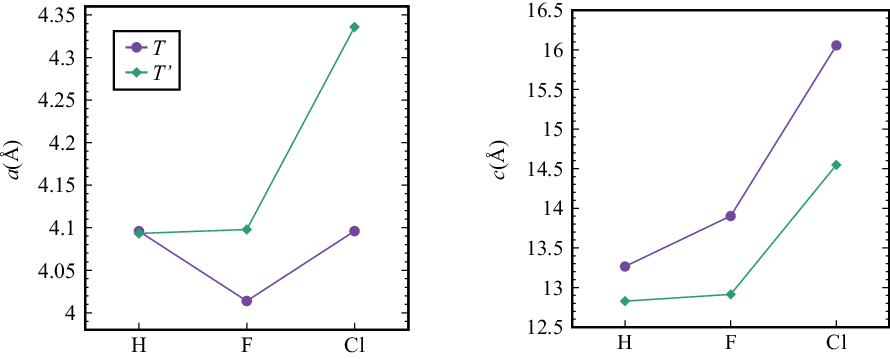}
\caption{Optimized lattice constants of Sr$_2$PdO$_2 X_2$.\label{fig8}}
\end{figure}   
\begin{figure}
\includegraphics[width=8cm]{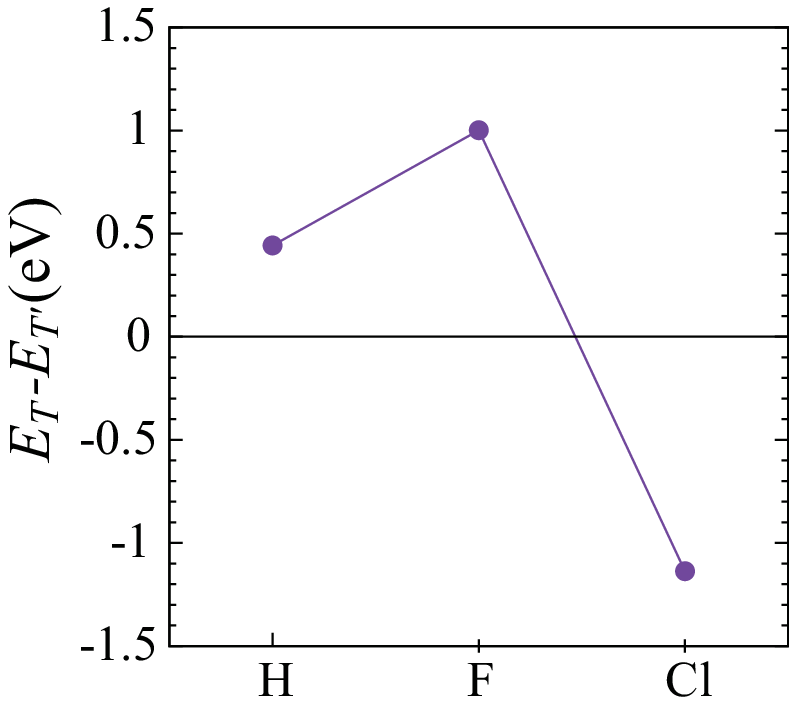}
\caption{Total energy difference between $T-$ and $T'-$ structures of Sr$_2$PdO$_2 X_2$.\label{fig9}}
\end{figure}

\begin{figure*}
\includegraphics[width=12cm]{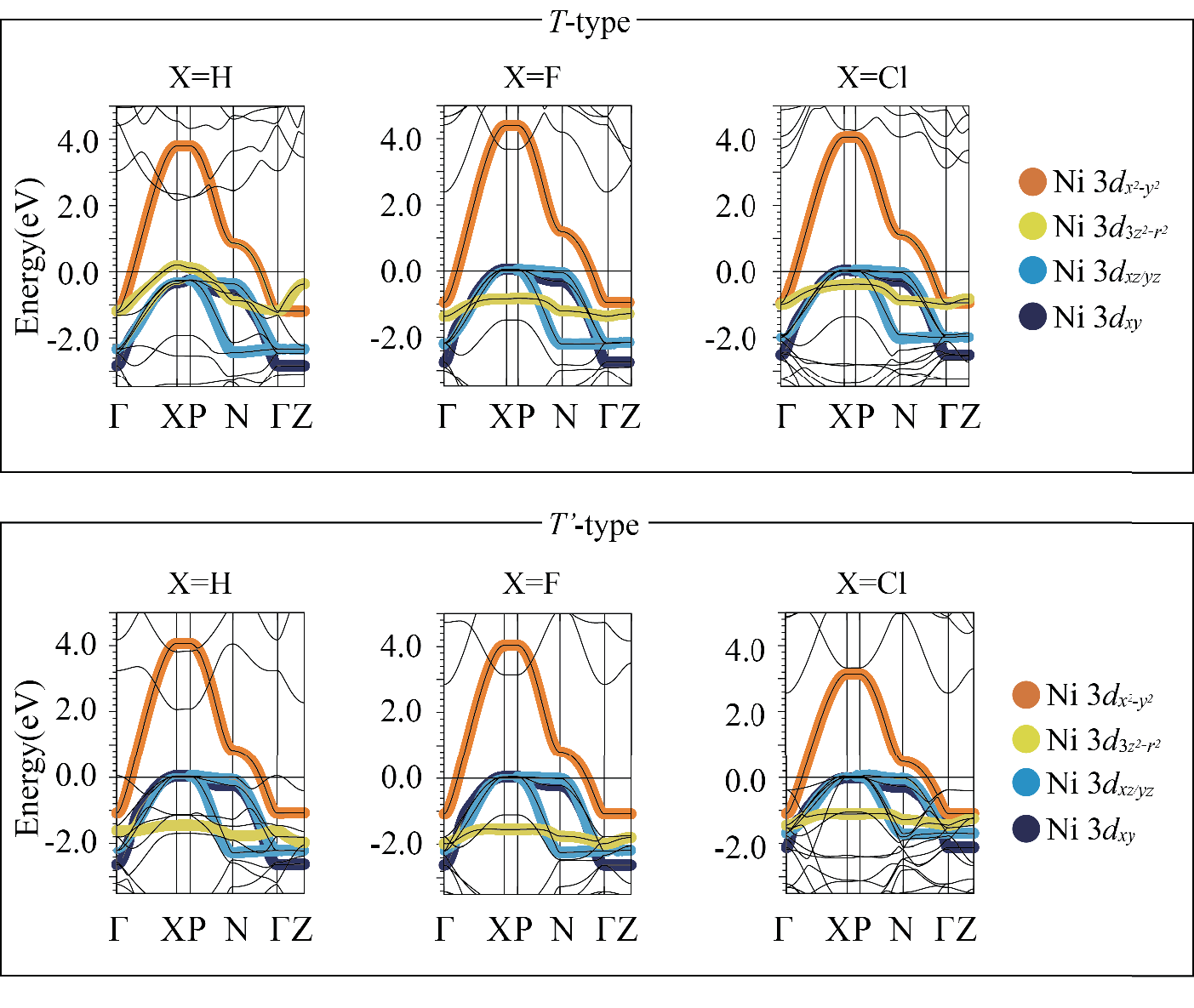}
\caption{Band structures of Sr$_2$PdO$_2 X_2$ for various $X$ and lattice types.\label{fig10}}
\end{figure*}   
\begin{figure}
\includegraphics[width=8cm]{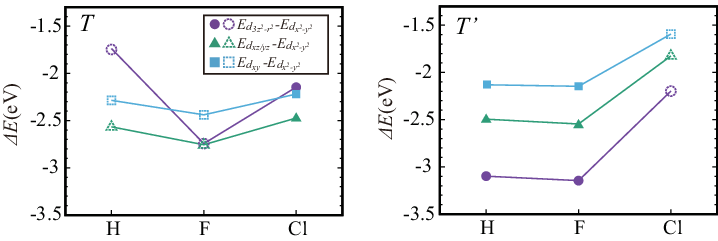}
\caption{The energy level offsets in Sr$_2$PdO$_2X_2$ between $d_{x^2-y^2}$ and other $d$ orbitals.\label{fig11}}
\end{figure}   
\begin{figure}
\includegraphics[width=8cm]{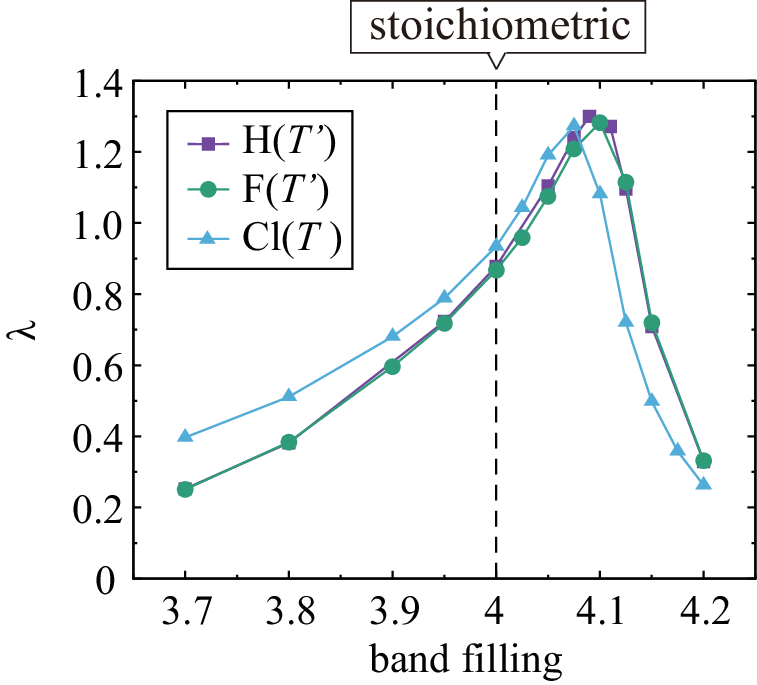}
\caption{The eigenvalue of the linearized Eliashberg equation at $T=0.01$eV against the band filling for the optimized crystal structures of Sr$_2$PdO$_2X_2$. Note that $\lambda$ exceeding unity indicates that $T_c$ is higher than $T=0.01$eV.\label{fig12}}
\end{figure}
\begin{figure*}
\includegraphics[width=16cm]{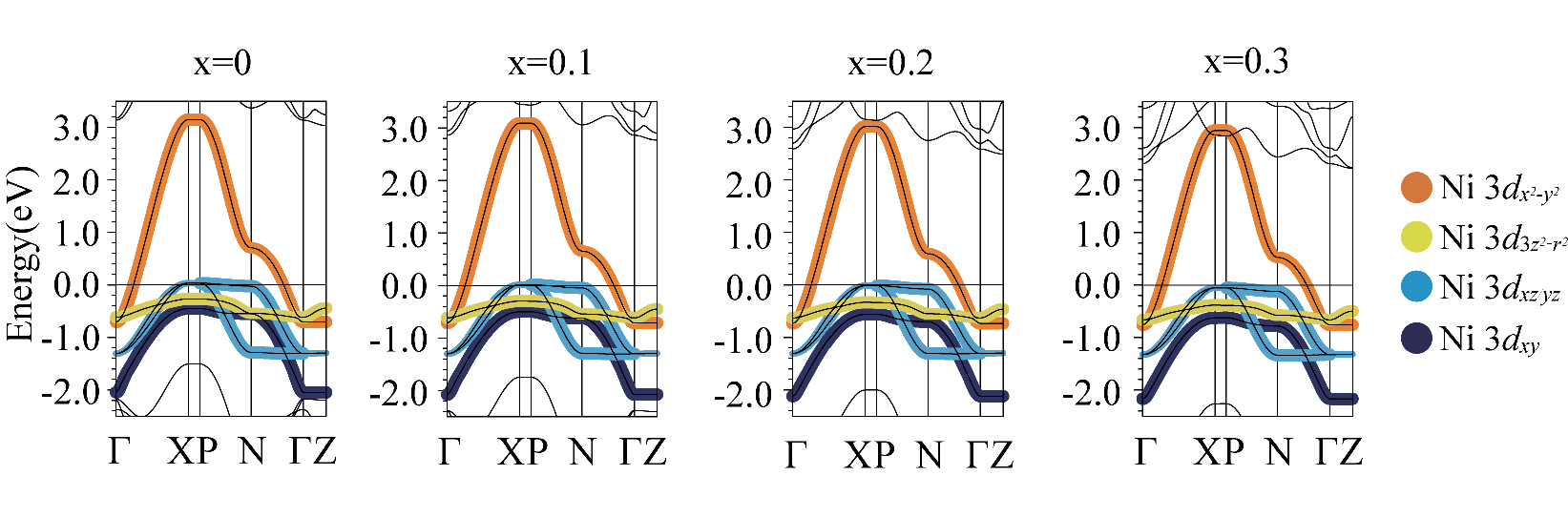}
\caption{Band structures of Ca$_{2-x}$La$_x$NiO$_2$Cl$_2$ obtained for various $x$ within the virtual crystal approximation.\label{fig13}}
\end{figure*}   
\begin{figure}
\includegraphics[width=8cm]{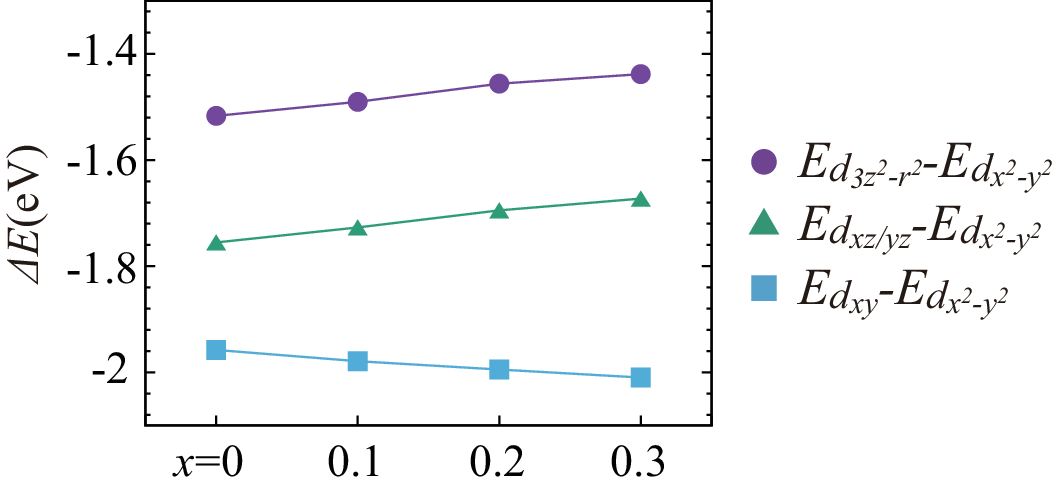}
\caption{$\Delta E$ for Ca$_{2-x}$La$_x$NiO$_2$Cl$_2$ obtained within the virtual crystal approximation and plotted against $x$. \label{fig14}}
\end{figure}   
\begin{figure}[h]
\includegraphics[width=8cm]{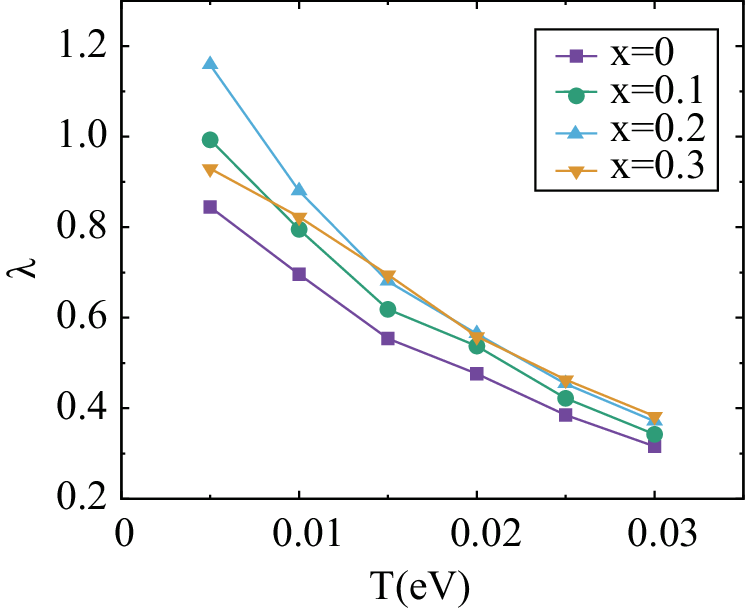}
\caption{Eigenvalue of the Eliashberg equation for Ca$_{2-x}$La$_x$NiO$_2$Cl$_2$ obtained within the virtual crystal approximation and plotted against temperature.\label{fig15}}
\end{figure}

\section{Methods}
We perform structural optimization using the PBE-GGA exchange-correlation functional\cite{PBE-GGA} and the projector augmented wave method\cite{Kresse}. For this purpose, we
use Vienna ab initio Simulation Package (VASP)\cite{VASP1,VASP2,VASP3,VASP4}. $12\times 12\times 12$ $k$ mesh and a plane-wave cutoff energy of 550 eV are used. Structural optimization is performed untile the Hellmann-Feynmann force becomes less than 0.01 eV \AA$^{-1}$ for every atom.
After the structural optimization, we perform first-principles band-structure calculation using WIEN2k code\cite{Wien2k}. We adopt RKmax=7 (6 for oxy-hydrides), and take $12\times 12\times 12$ $k$ mesh in the self-consistent-field calculations. From the calculated band structures, we extract the Wannier functions\cite{Marzari,Souza} of five Ni $3d$ orbitals using the Wien2Wannier\cite{w2w} and Wannier90\cite{Wannier90} codes. Throughout the study, the spin-orbit coupling is neglected.

We analyze superconductivity based on the five-orbital Hubbard models. We assume on-site intra- and inter-orbital interactions, $U$, $U'$, $J$ and $J'$, and the many-body study of this model is performed within the FLEX\cite{Bickers}. We mainly adopt $U=4$eV, $J=J'=U/10$, $U'=U-2J$. We obtain the renormalized Green's function by solving the Dyson's equation in a self-consistent calculation.  
The obtained Green's function and the pairing interaction mediated mainly by spin-fluctuations are plugged into the linearized Eliashberg equation. 
Since the eigenvalue $\lambda$ of the equation reaches unity at $T=T_c$, here we adopt $\lambda$, obtained at a fixed temperature of $T=0.01$eV, to measure how close the system is to superconductivity. The eigenfunction of the Eliashberg equation will be called the gap function. 
In the FLEX calculation, $16\times 16\times 4$ $(k_x,k_y,k_z)$ mesh and 2048 Matsubara frequencies are taken.

\section{Results}
\subsection{Pressure effects on Sr$_2$NiO$_2$Cl$_2$} 
The lattice constants obtained by optimizing the crystal structure under pressure are shown in Fig.\ref{fig3}. Both the in-plane ($a$) and out-of-plane ($c$) lattice constants monotonically decrease as expected. Here, the reduction of the in-plane lattice constant is especially important as this is expected to push up the $d_{x^2-y^2}$ energy level and hence enhance $\Delta E$. The band structures under pressure and the pressure dependence of $\Delta E$ are shown in Figs.\ref{fig4} and \ref{fig5}, respectively. As expected, the on-site energy differences between $d_{x^2-y^2}$ and other orbitals increase upon applying pressure.

In Fig.\ref{fig6}, we show the band filling dependence of the eigenvalue of the Eliashberg equation $\lambda$ for the three pressure cases. As already studied in Ref.\cite{Kitamine}, the maximum value of $\lambda$ is rather small at ambient pressure. As the pressure is increased, the maximum value of $\lambda$ is enhanced, and the maximum value is reached for smaller electron doping (i.e., band filling-4). This variance of $\lambda$ against the band filling under pressure is in fact reminiscent of that of  Ca$_2$NiO$_2$Cl$_2$ obtained in Ref.\cite{Kitamine}.

The gap function for the case that gives the largest $\lambda$ is shown in Fig.\ref{fig7}. As studied in Ref.\cite{Kitamine}, the gap function changes its sign between $d_{x^2-y^2}$ and other orbitals.

In total, our study shows that applying pressure to Sr$_2$NiO$_2$Cl$_2$ may lead to a possible high $T_c$ superconductivity, but we have to be cautious about several points. $\lambda$ values are not so large at the stoichiometric composition (band filling =4) even at the highest pressure, so in reality, electron doping might be necessary. Another point to be noted is that a low spin state is assumed in our study, while experiments  suggest that Sr$_2$NiO$_2$Cl$_2$ takes a high spin state\cite{Tsujimoto}. Our expectation is that reducing the lattice constant and increasing $\Delta E$ would result in a low spin state, as known empirically\cite{Kageyama}, but this point has to be examined theoretically in the future. 

\subsection{Sr$_2$PdO$_2 X_2$ ($X=$ H, F, Cl)}
In this section, we consider a 4d analogue of Sr$_2$NiO$_2 X_2$, namely, Sr$_2$PdO$_2 X_2$. Our main aim here is to investigate how $\Delta E$ is affected by substituting Ni by Pd. As was done in Ref.\cite{Kitamine}, we start with comparing the total energy of $T$ and $T'$ lattice structures. The optimized lattice constants and the energy difference between the $T$ and $T'$ structures are depicted in Figs.\ref{fig8} and \ref{fig9}, respectively. For $X=$Cl, the $T$ structure is found to have lower energy, while the $T'$ structure has lower energy for $X=$F and H. The latter may be consistent with the experimental observation that some oxy-fluroides such as Ba$_2$PdO$_2$F$_2$\cite{Baikie} and Sr$_2$CuO$_2$F$_2$\cite{Kissick} actually take the $T'$ structure.

The band structures for all the palladates considered are displayed in Fig.\ref{fig10}. The corresponding on-site energy level offsets between $d_{x^2-y^2}$ and other orbitals are shown in Fig.\ref{fig11}. Interestingly, $\Delta E$'s turn out to be significantly large, and especially for the lattice structure ($T$ or $T'$) with lower energy, they are even larger than those of Sr$_2$NiO$_2$Cl$_2$  under pressure. The origin of this large $\Delta E$ is likely because the spread of the 4d orbitals is larger than that of the 3d orbitals, resulting in an effective application of a certain kind of chemical  pressure. Large $\Delta E$ found in Sr$_2$NiO$_2$Cl$_2$ is consistent with the experimental observation of low spin state in Ba$_2$NiO$_2$Cl$_2$\cite{Tsujimoto}.

Due to the increase of $\Delta E$, the eigenvalue of the Eliashberg equation $\lambda$ is significantly enhanced compared to the case of the nickelates, as shown in Fig.\ref{fig12}. In fact, the maximum value of $\lambda$ exceeds unity even at the present temperature of  $T=0.01$eV. This result suggests that $T_c$ is higher than the models of the high-$T_c$ cuprates\cite{SakakibaraNi}, although we do have to be somewhat cautious about the interpretation of this result because 4d orbitals  are less localized compared to 3d, so that adopting Hubbard type models with only on-site interactions is less justified.  

\subsection{The effect of band structure variation with doping}
We have seen that certain amount of electron doping (i.e., band filling larger than 4) is required in order to optimize superconductivity in the nickelates proposed here. On the other hand, the band filling is varied within the rigid band approximation in our calculation. In reality, the partial substitution of the elements required for doping modifies  the band structure from that of the stoichiometric composition. Another concern of our analysis so far is that we have only considered the eigenvalue $\lambda$ of the Eliashberg equation at $T=0.01$eV as a measure for the superconducting transition temperature $T_c$. It remains somewhat uncertain whether the actual $T_c$, where $\lambda(T_c)=1$, corresponds to the magnitude of $\lambda(T=0.01{\rm eV})$. In this section, we check these issues.

We perform the band structure calculation of Ca$_{2-x}$La$_x$NiO$_2$Cl$_2$ within the virtual crystal approximation. The band structures for $x=0,0.1,0.2$ and 0.3 are shown in Fig.\ref{fig13}. The corresponding level offsets $\Delta E$ are no longer constants and now vary against $x$, as depicted in Fig.\ref{fig14}. We can see, however, that the variance is not large. We plot $\lambda$ against the temperature for these values of $x$ in Fig.\ref{fig15}. It can be seen that the order of $T_c$ among the four values of $x$ coincides with that of $\lambda(T=0.01{\rm}eV)$, except between $x=0.1$ and $x=0.3$, where $\lambda$ of the latter case tends to saturate upon lowering the temperature. Such a tendency of $\lambda$ against the temperature is seen when the incipient band is too far away from the Fermi level. In any case, $T_c$ defined by $\lambda(T_c)=1$ is maximized between $x=0.1$ and $x=0.2$, which is consistent with our rigid band analysis using $\lambda(T=0.01{\rm}eV)$.

\section{Conclusions}
In this paper, we have studied possible ways of realizing unconventional superconductivity in mixed-anion nickelates and palladates with $d^{8+\delta}$ electron configuration. The underlying idea is based on the mathematical equivalency between the  two-orbital Hubbard model and the bilayer Hubbard model, where $\Delta E$ in the former corresponds to $2t_\perp$ in the latter.\cite{Shinaoka}.  Although mixed-anion nickelates with large $\Delta E$ such as Ca$_2$NiO$_2$Cl$_2$ were proposed as good candidates in our previous study\cite{Kitamine}, there can be difficulties in synthesizing these materials. Here, we have shown that an alternative way of realizing a similar situation is to apply pressure to  Sr$_2$NiO$_2$Cl$_2$, which enlarges $\Delta E$ through enhanced crystal field effect. Going over to a 4d analogue, Sr$_2$PdO$_2$Cl$_2$ turns out to be another possibility due to larger $\Delta E$ compared to the nickelates.  Most of our studies assume a rigid band in showing that a certain amount of electron doping should enhance superconductivity. In order to check the effect of the band variation upon electron doping, we have studied Ca$_{2-x}$La$_x$NiO$_2$Cl$_2$ within the virtual crystal approximation, which has given consistent results with the rigid band  analysis. 


\begin{acknowledgments}
This study has been supported by JSPS KAKENHI Grant No. JP22K04907 (K. K.).
The computing resource is supported by 
the supercomputer system (system-B) in the Institute for Solid State Physics, the University of Tokyo.
\end{acknowledgments}

\bibliography{srnioclpress2}

\end{document}